\documentstyle[aps,epsf]{revtex}  

\begin{document}        

\baselineskip 14pt
\title{Global Analysis of Solar Neutrino Data}
\author{P.I. Krastev}
\address{University of Wisconsin-Madison}
%
\maketitle              

\begin{abstract}        
The data from all solar neutrino experiments presented at the Neutrino
98 conference has been analyzed and a brief summary of the status of
the two most popular neutrino solutions of the solar neutrino problem,
the MSW mechanism and vacuum oscillations, is presented here.  The
inclusion of the recoil-electron data and of the zenith angle
distribution of events in the SuperKamiokande detector, together with
the routine analysis of the rates, into a global $\chi^2$ fit impose
stringent constraints on these neutrino oscillation scenarios.
\end{abstract}   	

\section{Introduction}               

Analyses of solar neutrino data in the past have focused mostly on
fitting the observed rates from the four pioneering experiments
\cite{chlorine}+\cite{kamiokande}+\cite{SAGE}+\cite{GALLEX} and
finding the allowed regions in the relevant parameter space. The
recoil electron data, as well as the zenith angle dependence of events
in the Kamiokande experiment, have been also used separately to
constrain, independently of the predicted $^8{\rm B}$ flux, neutrino
oscillation parameters.  The latest data from SuperKamiokande
\cite{neutrino98}+\cite{superkamiokande} is remarkable and
unprecedented in it's precision.  The time has come for a global
approach in which data from the measurement of the rates in all solar
neutrino experiments is combined in a single fit with data from the
recoil electron spectrum and zenith angle distribution of events in
this detector. This procedure results in more stringent constraints on
the neutrino parameters, as well as a better understanding of the
status of the neutrino oscillation solutions to the solar neutrino
problem.

\section{Rates Only}

\begin{figure}[ht]	
\centerline{\epsfxsize 5.0 truein \epsfbox{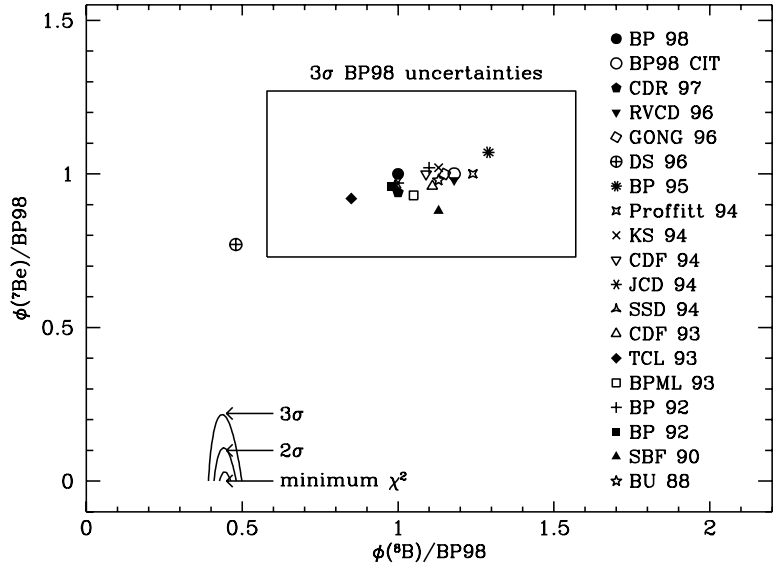}}   
\vskip -6.0 cm
\caption[]{
\label{Smodels}
\small Standard solar model predictions (marked by various symbols)
versus a solar model independent fit (three ellipses) to all
experiments assuming arbitrary neutrino flux normalization with no
energy distortion.}
\end{figure}

The analysis of the rates from all experiments is a first step in the
global approach. The rates are probably the most robust piece of
data. The experimental results together with the predictions from the
standard solar model (SSM) \cite{BP98} are summarized in
Table.\ref{datarates}. Different groups have shown \cite{SMod} that
the analysis of the rates from Homestake, Kamiokande, GALLEX and SAGE
leave little hope for an astrophysical solution. The primary reason is
that the combined solar neutrino data appears to indicate that the
beryllium neutrinos are missing \cite{BB}.  This important result is
reconfirmed by the analysis of the latest solar neutrino data
including the SuperKamiokande result.  Allowing for all fluxes to vary
as free parameters, but keeping the shape of the spectrum from
individual sources unchanged, provides a poor fit to the measured
rates from all solar neutrino experiments. Any such solution is ruled
out at 99 \% C.L. The resulting 1-, 2-, and 3-$\sigma$ contours in the
space of $^8{\rm B}$-$^7{\rm Be}$ flux are shown in Fig.\ref{Smodels}
together with the predictions of 19 solar models (see \cite{BKS} for
details).

\begin{table}
\caption[]{Solar neutrino data and SSM predictions used in the
analysis. The units in the second and third columns are SNU for all of
the experiments except Kamiokande and SuperKamiokande, for which the
measured (predicted) $^8{\rm B}$ flux at the earth is given,
correspondingly above 7.5 MeV and 6.5 MeV, in units of ${\rm 10^6
cm^{-2}s^{-1}}$. The ratios of the measured values to the
corresponding predictions in the Bahcall-Pinsonneault SSM of
Ref. \protect\cite{BP98} are given in the fourth columnwhere only the
experimental errors are included.  The results cited for the
Kamiokande and SuperKamiokande experiments assume that the shape of
the ${\rm ^8B}$ neutrino spectrum is not affected by physics beyond
the standard electroweak model.\label{datarates} }
\begin{tabular}{l c c c c c}
Experiment & Result & Theory & Result/Theory & Reference \\
\hline

Homestake & $2.56 \pm 0.16 \pm 0.14$ &
$7.7^{+1.2}_{-1.0}$ & $0.33 \pm 0.028$ 
&\cite{chlorine}\\

Kamiokande & $2.80 ~\pm 0.19 ~\pm 0.33$ & $5.15 ~^{+1.0}_{-0.7}$ 
& $0.54 \pm 0.07$ &\cite{kamiokande}\\ 

SAGE & $70.3 ~{}^{+8}_{-7.7}$ & $129^{+8}_{-6}$&
$0.54 \pm 0.06$ &\cite{SAGE} \\

GALLEX & $76.4 ~\pm 6.3 ~{}^{+4.5}_{-4.9}$ & $129^{+8}_{-6}$&
$0.59 \pm 0.06$ &\cite{GALLEX}\\

SuperKamiokande & $2.37 ~{}^{+0.06}_{-0.05} ~{}^{+0.09}_{-0.07}$ 
& $5.15 ~^{+1.0}_{-0.7}$& $0.46 \pm 0.020$ 
&\cite{superkamiokande}\\ 

\end{tabular}
\end{table}

\begin{figure}[ht]	
\centerline{\epsfxsize 5.0 truein \epsfbox{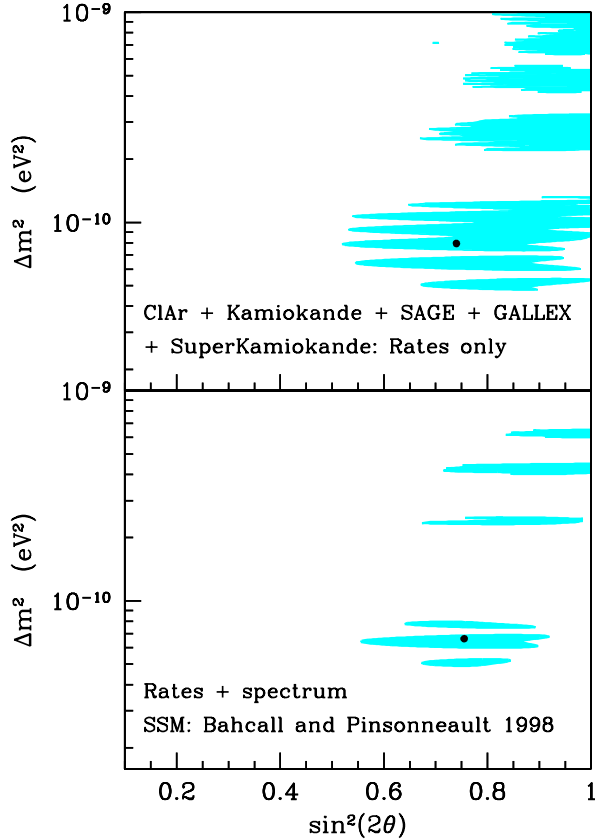}}   
\vskip -.4 cm
\caption[]{
\label{vacuum}
\small Allowed regions in parameter space for neutrino oscillations in
vacuum. The top panel describes the allowed regions obtained by
fitting only the rates from all solar neutrino experiments. The bottom
panel shows the result of including the information from the
recoil-electron spectrum in SuperKamiokande.}
\end{figure}

It has been known for quite some time (see \cite{BKS} and references
therein) that since different experiments measure the solar neutrino
flux from different sources (pp, $^7{\rm Be}$, $^8{\rm B}$, etc), and
since the measured rates are not the same, the analysis of the rates
only limits the allowed regions to just a few small areas in parameter
space ($\Delta {\rm m}^2$ - $\sin^22\theta$) assuming either MSW
transitions or vacuum oscillations as solutions to the solar neutrino
problem. The MSW mechanism gives an excellent fit to the data from the
rates only. The allowed regions are given in the upper panel of
Fig.\ref{MSW}. The minimum $\chi^2$ is 1.7. Assuming 2 degrees of
freedom (d.o.f.) this corresponds to a 57 \% C.L. The vacuum
oscillation solution provides a somewhat worse fit: minimum $\chi^2$ =
4.3 (88 \% C.L.). The allowed regions are shown in the upper panel of
Fig.\ref{vacuum}. Although the MSW mechanism provides a better fit to
the data, vacuum oscillations are still a viable solution to the solar
neutrino problem.

\section{Spectral Distortions}

\begin{figure}[ht]	
\centerline{\epsfxsize 5.0 truein \epsfbox{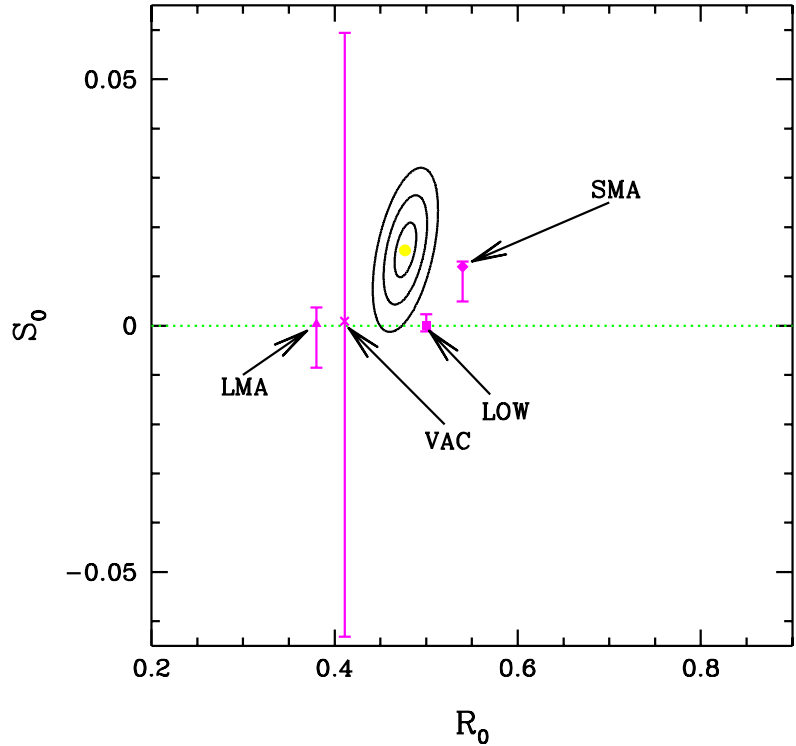}}   
\vskip -.2 cm
\caption[]{
\label{Ellips}
\small The results of a fit of a straight line parameterized by an
intercept (${\rm R}_0$) and slope (${\rm S}_0$) to the ratio of the
measured by SuperKamiokande recoil-electron spectrum to the expected
undistorted spectrum. The three ellipses are the 1-, 2- and 3-$\sigma$
contours representing the fit. Also shown are the predicted values for
the slope and intercept parameters in four neutrino oscillation
solutions together with the respective error bars on ${\rm R}_0$. The
corresponding error bars on ${\rm S}_0$ are not shown but they are
wide enough to cross at least some of the ellipses.}
\end{figure}

Neutrino physics solutions predict, for a wide and plausible range of
parameters, different suppression factors for different sources, thus
leading to an energy dependent suppression of the solar neutrino
spectrum. Remarkably, these solutions in general also predict
distortions of some sort in the spectra of individual neutrino
sources. Kamiokande and SuperKamiokande are the first detectors able
to test this prediction for the boron neutrinos. The error bars of the
Kamiokande data were not small enough for a definite conclusion to be
reached about this important point. The data were consistent with no
distortion but could not test the small distortions predicted within
some of the parameter space, favored by the analysis of the rates. The
data from SuperKamiokande after only 504 days of operation, are
significantly more accurate. This is a result of both the superior
statistics and truly heroic efforts on the part of the collaboration
to reduce the systematic errors, including a calibration with a
``portable'' linac. The comparison of the measured spectrum of
recoil-electrons with the one expected in the absence of oscillations
reveals a distortion, namely the spectrum of recoil electrons is
incompatible with an undistorted one at the 95 \% C.L.  The result of
fitting a straight line to the recoil-electron data in SuperKamiokande
is given in Fig.\ref{Ellips}. together with the range of values of the
two parameters (slope (${\rm S}_0$) and intercept (${\rm R}_0$))
obtained when the neutrino oscillation parameters are varied within
the corresponding allowed regions. Although the best fit points to the
rates predict slope and intercept outside the 99 \% C.L. ellipse, the
experimental errors and theoretical uncertainties are still not small
enough to rule out any of these solutions on the basis of the
recoil-electron spectrum only.

\section{Rates and Recoil-Electron Spectrum}

When the data from the rates measured in all solar neutrino
experiments are analyzed together with the data from the spectrum of
recoil-electrons in SuperKamiokande, the combined $\chi^2$ fit becomes
worse than in either of the two separate fits. The MSW solution
provides a minimum $\chi^2$ of 26.5 \% corresponding to a 93 \%
C.L. (17 d.o.f.). The 99 \% C.L. allowed regions are shown in the
middle panel of Fig.\ref{MSW}. The vacuum oscillations solution gives
a slightly better fit: $\chi^2_{\rm min}$ = 22.8 (84 \% C.L.). The
corresponding 99 \% C.L. allowed regions are shown in Fig.\ref{MSW}.
The primary reason for the not particularly good fit is the mismatch
between the best fit points in the two separate fits. Since the fit to
the rates-only is considerably better than the fit to the spectrum,
the first is driving the combined fit. Vacuum oscillations fit better
the high-energy part of the spectrum near the end point of the $^8{\rm
B}$ spectrum. This is explained by the steeper increase of the
survival probability with energy that vacuum oscillations can provide,
if the parameters $\Delta {\rm m}^2$ and $\sin^22\theta$ are chosen
properly. Future refinements of the measurement of the spectrum can
change the situation. A confirmation of the excess of events above 13
MeV would favor vacuum oscillations or a higher flux of hep-neutrinos
(see section \ref{sectionhep}). Note also that the more recently
published 708 days of data \cite{smy} show a smaller distortion of the
spectrum which almost certainly will improve the overall fit.

\section{Rates, Electron Spectrum and Zenith Angle Distribution}

The inclusion of the zenith angle distribution data from
SuperKamiokande is a very important step in the global fit. A
significant virtue of the SuperKamiokande detector is its
directionality.  The direction of the electrons scattered from solar
neutrinos points backwards to the sun. The zenith angle dependence of
these events must have a particular shape that can be accurately
calculated. Since for a wide range of masses and mixing angles the MSW
effect predicts a significant regeneration of electron neutrinos, and
therefore an enhanced signal at night, while neutrinos pass through
the earth, the zenith angle distribution should be distorted in a
specific and calculable way \cite{brighter}. The comparison of the
measured zenith angle distribution with the one predicted in the MSW
mechanism limits the allowed parameter space. A large region is ruled
out independently of the absolute flux of boron neutrinos (see
Ref.\cite{BKS} for references and details).

\begin{figure}[ht]	
\centerline{\epsfxsize 5.0 truein \epsfbox{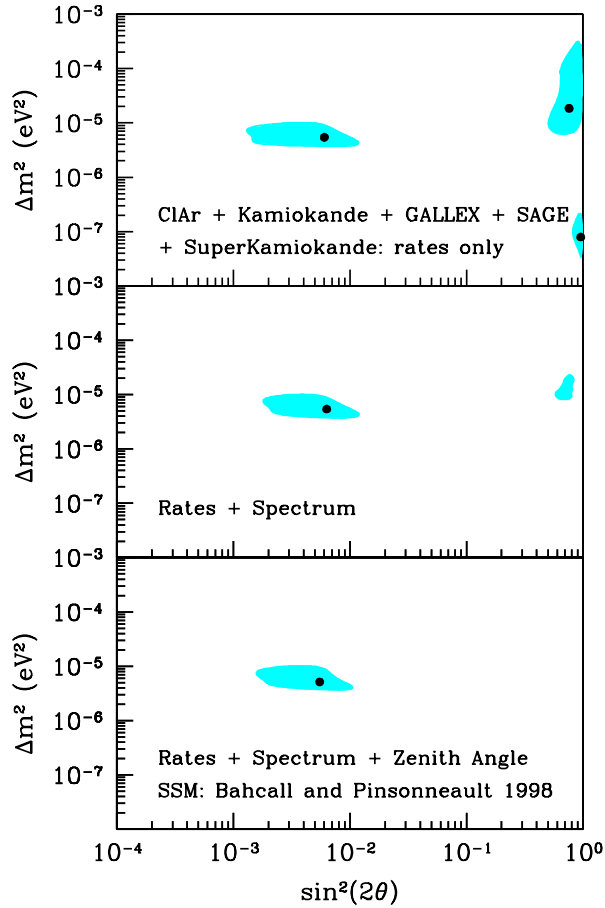}}   
\caption[]{
\label{MSW}
\small Allowed regions in parameter space for resonant neutrino
transitions in matter (MSW effect). The top panel is obtained by
fitting only the rates from the five solar neutrino experiments. The
middle panel presents the result of including the recoil-electron
spectrum in SuperKamiokande and the bottom panel is the final allowed
region obtained by including the zenith angular distribution of recoil
electrons in the global fit.}
\end{figure}

A common misconception sometimes comes up when discussing either the
spectral distortions, or the zenith angle distribution of events in
water-Cherenkov detectors. It is occasionally being used as an
argument against neutrino physics solutions to the solar neutrino
problem. Since none of these distortions has been observed, some
researchers tend to make the wrong conclusion that either MSW or
vacuum oscillations, or both, are either ruled out or are at least in
a serious trouble. In fact both the MSW mechanism and vacuum
oscillations predict a wide range of spectral distortions. The MSW
effect predicts quite strong but also negligible small zenith angle
distortions. The data is just not sufficient yet to rule the predicted
distortions of either type, if the global best fit points are the ones
chosen by nature. The predicted distortions corresponding to the best
fit points can be tested at the 3-$\sigma$ level with a data set at
least three times larger than the present data, assuming the
systematic errors can be reduced correspondingly (see
Ref.\cite{brighter} and \cite{BKL} for details).

The global fit to the rates, recoil-electron spectrum and zenith angle
distribution is not particularly good. The minimum $\chi^2$ for MSW is
37.2 (93 \% C.L. for 26 d.o.f). The resulting 99 \% C.L. allowed
region is shown in the lower panel in Fig.\ref{MSW}. The two large mixing
angle solutions are ruled out at this C.L.  Vacuum oscillation give a
minimum $\chi^2 = 28.4$ (94 \% C.L., 18 d.o.f.). The C.L. for the MSW
and vacuum oscillation solutions are very close and therefore none of
these solutions can be claimed to be better than the other.  In future
the fit can go in either direction as a result of refined measurements
of the spectrum and/or of the zenith angle distribution of events.

\section{Hep neutrinos}
\label{sectionhep}
\begin{figure}[ht]	
\centerline{\epsfxsize 4.0 truein \epsfbox{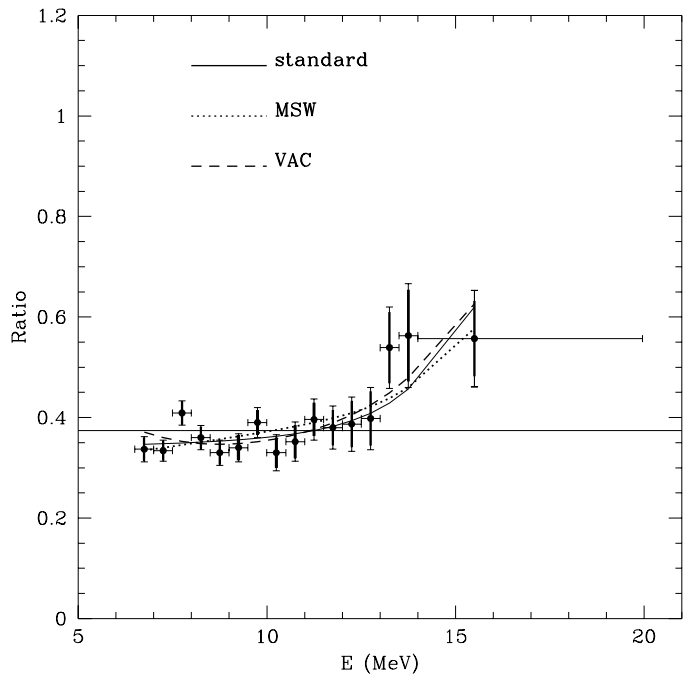}}   
\vskip -.2 cm
\caption[]{
\label{hep}
\small The recoil electron spectrum measured by the SuperKamiokande
collaboration (504 days of data) is fitted with a hep-neutrino flux 26
times larger than the one in the standard solar model. The rise of the
predicted spectrum at the end-point is due to hep neutrinos}
\end{figure}

In the standard solar model the flux of (hep) neutrinos from the
$^3{\rm He}$(p, ${\rm e}^+$$\nu_{\rm e}$)$^4{\rm He}$ reaction is
about 2000 times smaller than the flux of $^8{\rm B}$ neutrinos. This
result is due to the very small calculated cross-section for this
reaction.  Experientally it is very difficult, if not impossible, to
measure the cross-section of the above reaction in a laboratory on
earth. However, an amusing feature of this reaction is that the
hep-neutrinos have a higher endpoint (18.8 MeV) than the $^8{\rm B}$
(16 MeV) neutrinos. This could explain the apparent sharp rise of the
recoil electron spectrum above 13 MeV (which in fact was indicated
also in the old Kamiokande data), if the flux of hep-neutrinos is
higher by a factor of about 20-30 \cite{hep} than in the standard
solar model.  The predicted spectrum assuming $\Delta {\rm m}^2$ and
$\sin^22\theta$ equal to the best fit points from a global $\chi^2$
analysis including rates, electron spectrum, zenith angle distribution
and an arbitrary hep-neutrino flux, is shown in Fig.\ref{hep}.  By
accurately measuring the spectrum above 13 MeV the SuperKamiokande
collaboration can test this interesting possibility.

\section{Conclusions}

With the latest data from the SuperKamiokande experiment a global
analysis of all available solar neutrino data becomes a necessity.
The first results of such an analysis are summarized above for
oscillations into active neutrinos. The correspoding results for
sterile neutrinos can be found in \cite{BKS}. Results presented by the
SuperKamiokande collaboration at this meeting \cite{smy} show smaller
excess of events near the end-point of the $^8{\rm B}$ spectrum. The
study of the corresponding implications for neutrino oscillation
solutions of the solar neutrino problem is underway. The results are
likely to change quantitatively by a small amount but are unlikely to
change qualitatively the whole picture. Further studies with detectors
like SNO \cite{SNO}, Icarus \cite{icarus}, Borexino \cite{Borexino}
and Hellaz \cite{hellaz} remain indispensible.

\section{Acknowledgements}

I am grateful to J. Bahcall and A. Smirnov for a stimulating 
collaboration. This work has been supported by an NSF grant
PHY-9605140.

\end{document}